\documentclass{eas}
\usepackage{graphicx}
\begin{document}

%%-----------------------------
%%      the top matter
%%-----------------------------
\title{On the Cepheid Metallicity Dichotomy} 
\author{G\'eza Kov\'acs}
\address{Konkoly Observatory, Budapest, Hungary}
%\author{...}\address{...}
%
%
\begin{abstract}
With the aid of stellar atmosphere models, we derive iron 
abundances [Fe/H] from the OGLE $B$, $V$, $I$ photometry 
on the Magellanic Cloud Cepheids. We show that in both 
clouds the average metallicities of the first overtone 
variables are lower than those of the fundamental ones 
(by $\sim 0.2$ and $\sim 0.3$~dex in the LMC and SMC, 
respectively). Consequently, there is a correlation between 
the overall [Fe/H] and luminosity; the lower luminosity 
stars tend to be also of lower metallicity. These metallicity 
dependencies are concordant with the ones derived for the 
two types of double-mode Cepheids from pulsation theory. 
Yet another support of this dichotomy comes from the evolution 
theories that require lower metallicities for blue-looping 
low-luminosity stars than for high-luminosity ones. We also 
comment on the possibility of using period-luminosity-color 
relations to derive more accurate metallicities. 
\end{abstract}
\maketitle

%%-----------------------------
%%      your text
%%-----------------------------

%
%%%%%%%%%%%%%%%%
%  SECTION 1
%%%%%%%%%%%%%%%%
%
\section{Introduction}
It is well known that chemical composition plays a very important 
role in stellar evolution and in various methods of stellar 
parameter computations, most remarkably in asteroseismological 
investigations. From the point of view of chemical composition, 
stellar clusters and certain populations of galaxies have been 
treated quite often as chemically homogeneous ensembles. However, 
accumulating abundance measurements show that there are chemical 
inhomogeneities even in globular clusters (e.g., Yong \& Grundahl 
\cite{yong08}), where the uniform composition thought to be 
well-justified on the basis of common origin of the constituting 
objects. It has also been evident during the recent years -- 
mostly due to the systematic spectroscopic surveys of Andrievsky 
and coworkers -- that there is a considerable metallicity spread 
among the Galactic Cepheids and that the observed metallicity 
correlates with the distance from the Galactic center 
(Luck, Kovtyukh, \& Andrievsky \cite{luck06}). The importance of 
the heavy element abundance in stellar pulsation has become especially 
clear after 1992, with the verification of Simon's hypothesis on 
the iron opacity bump (Simon \cite{simo82}; Rogers \& Iglesias 
\cite{roig92}). Perhaps the largest impact made by the opacity 
increase was the near elimination of the so-called `Cepheid mass 
discrepancy' (Moskalik, Buchler \& Marom \cite{mosk92}) and the 
excitation of the $\beta$~Cephei stars (Moskalik \& Dziembowski 
\cite{modz92}). In addition, based on these revised opacities, 
there are theoretical results which show that the consideration 
of metallicity spread is a must in various groups of Cepheids. 
Currently it has been shown that the observed periods of the 
first/second overtone double-mode Cepheids in the Magellanic 
Clouds can be fitted only if we assume that the metallicities 
of these low-luminosity stars are lower than their fundamental/first 
overtone counterparts (Kov\'acs \cite{koge06}). This observation 
supports the earlier result of Cordier, Goupil \& Lebreton 
(\cite{cord03}), suggesting that Cepheids cannot `blue-loop' 
at low-luminosities in the SMC, unless they have systematically 
lower metallicities than the higher luminosity ones. Stimulated 
by the above results, the goal of the present study is to use 
the photometric metallicities in the investigation of the 
metallicity spread among the Magellanic Cloud Cepheids. 

%
%%%%%%%%%%%%%%%%
%  SECTION 2
%%%%%%%%%%%%%%%%
%
\section{Metallicities from BVI colors}
Our method has already been described in D\'ek\'any et al. (\cite{deka08}) 
and tested on a sample of 21 fundamental mode Galactic field RR~Lyrae 
stars. Here we summarize the main steps of the method and present 
tests concerning the limits of applicability of the photometrically 
derived abundances. 

Assuming that the poorly known quantities, such as the turbulent 
velocity and convective mixing length are fixed to some generally 
acceptable value and the relative abundances are also given (e.g., 
scaled to the Sun), then stellar atmosphere models and basic 
pulsational relations yield the following set of equations: 

%
%%%%%%%%%%%%%%
%   EQ. (1)
%%%%%%%%%%%%%%
%
\begin{eqnarray}
B-V = F(T_{\rm eff}, g, {\rm [Fe/H]}) \hskip 2mm , \hskip 3mm 
V-I = G(T_{\rm eff}, g, {\rm [Fe/H]}) \hskip 2mm , \hskip 3mm 
g   = H(P) \hskip 2mm ,
\end{eqnarray}

\noindent
where the last equation is applicable in a wide range of parameters 
in the region of Cepheids (see Kov\'acs 2000). The functions $F$ 
and $G$ denote the dependence of the theoretical color indices 
on the physical parameters as given by the models of Castelli, 
Gratton \& Kurucz (\cite{cast97}). All models have solar-type 
heavy element distribution, microturbulent velocity of 2~km\,s$^{-1}$ 
and no convective overshooting. Quadratic interpolation among 
these models proved to be accurate enough to find the solution 
of set of Eqs.~(2.1), which is searched for by minimizing the 
following quantity: 

%
%%%%%%%%%%%%%%
%   EQ. (2)
%%%%%%%%%%%%%%
%
\begin{eqnarray}
{\cal D}^2 & = & 
\bigl [\Delta\log T_{\rm eff}\bigr ]^2 + 
\bigl[w_{\rm B-V}\Delta(B-V)\bigr]^2 +
\bigl[w_{\rm V-I}\Delta(V-I)\bigr]^2 
\hskip 2mm ,
\end{eqnarray}

\noindent
where $\Delta\log T_{\rm eff}$ is the difference between the 
$\log T_{\rm eff}$ values computed from the $B-V$ and $V-I$ 
colors, whereas $\Delta(B-V)$ and $\Delta(V-I)$ denote the 
differences between the observed values and the ones given by 
the stellar atmosphere model. The weights $w_{\rm B-V}$ and 
$w_{\rm V-I}$ are set equal to $0.33$ and $0.25$, respectively, 
following the proportionality of $\log T_{\rm eff}$ in the 
approximate linear expressions of Kov\'acs \& Walker (\cite{kowa99}). 
The last two terms in $\cal D$ increases the stability of the 
method against observational noise. In general, we accept 
solutions only with $\cal D$~$<0.001$.  

%
%>>>>>>>>>>
% FIG. 1
%>>>>>>>>>> 
%
\begin{figure}[h]
\includegraphics[width=55mm]{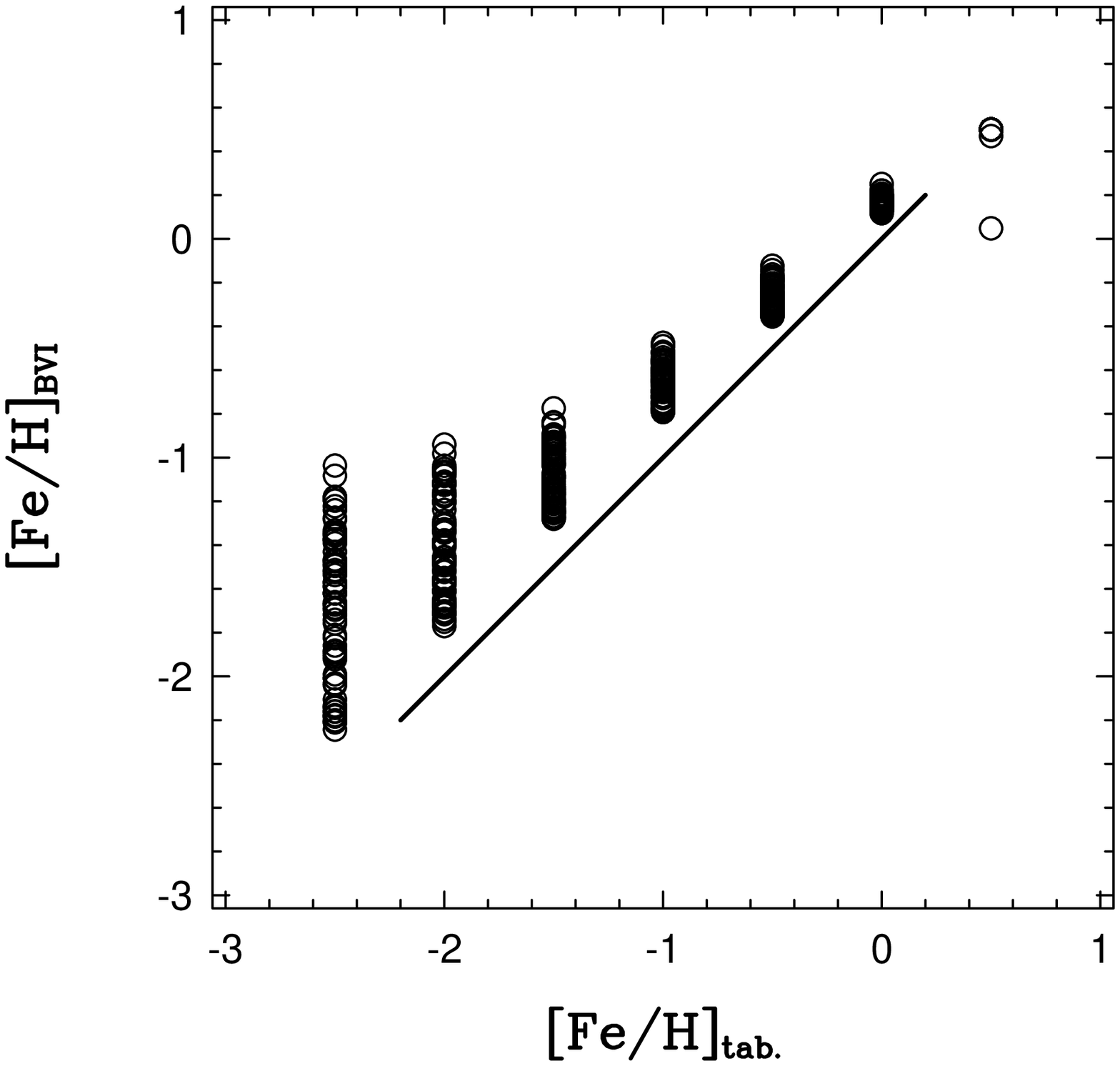}
\hskip 5mm
\includegraphics[width=55mm]{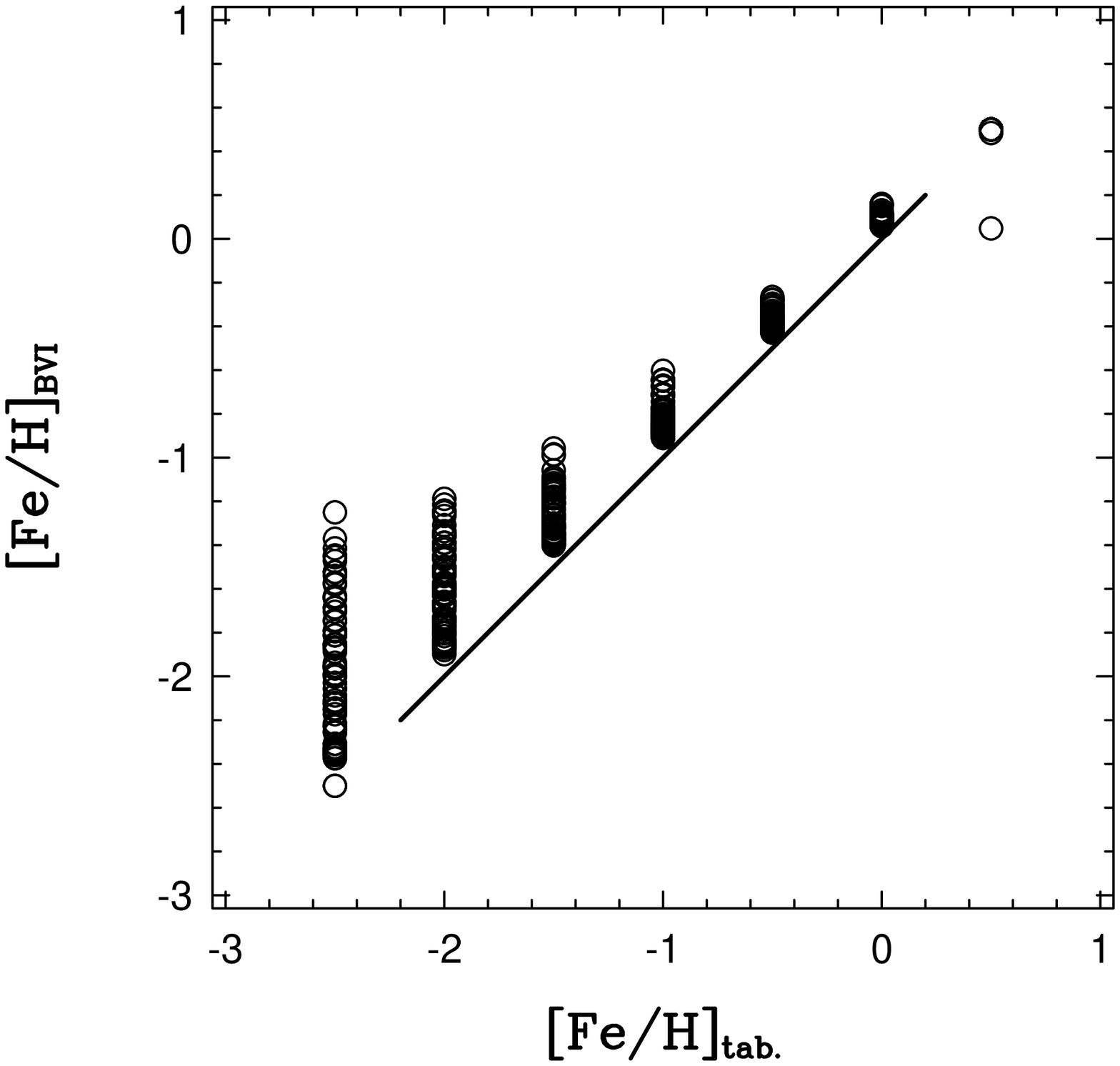}
\caption{{\it Left:} theoretical vs. computed photometric metallicities 
         by shifting the theoretical $V$ magnitudes by $\Delta V=-0.01$~mag 
	 and use this (with the other input parameters left at their 
	 theoretical values) in the computation of [Fe/H]$_{\rm BVI}$. 
	 {\it Right:} the same plot for the $B$ color with 
	 $\Delta B=+0.01$~mag.} 
\end{figure}

%
%&&&&&&&&&&&&
%  TABLE 1.
%&&&&&&&&&&&&
%
\begin{table}[h]
\caption[]{Systematic and random effects for the photometric [Fe/H] 
and $T_{\rm eff}$}
\label{}
\centering 
\begin{tabular}{lrcr}
\hline\hline
\multicolumn{4}{l}{\bf Systematic errors:}\\
Parameter           & shift    & $\Delta {\rm [Fe/H]}$ & $\Delta T_{\rm eff}$\\
\hline
$\Delta V$          & $+0.01$  & $-0.31\pm0.18$ & $-34\pm12$\\
                    & $-0.01$  & $+0.41\pm0.28$ & $+31\pm12$\\
$\Delta B$          & $+0.01$  & $+0.25\pm0.21$ & $\phantom{0} -3\pm\phantom{0} 7$\\
                    & $-0.01$  & $-0.17\pm0.12$ & $\phantom{0} +0\pm\phantom{0} 6$\\
$\Delta I$          & $+0.01$  & $+0.22\pm0.15$ & $+33\pm\phantom{0} 5$\\
                    & $-0.01$  & $-0.18\pm0.12$ & $-34\pm\phantom{0} 6$\\
$\Delta \log g$     & $+0.10$  & $-0.01\pm0.03$ & $\phantom{0} +8\pm\phantom{0} 7$\\
                    & $-0.10$  & $+0.01\pm0.04$ & $\phantom{0} -8\pm\phantom{0} 8$\\
\hline
\multicolumn{4}{l}{\bf Random errors:}\\
Parameter  & $\sigma$\ \ \ & $\sigma {\rm [Fe/H]}$ & $\sigma(T_{\rm eff})$ \\
\hline
$\sigma (B-V)=\sigma (V-I)$    & $0.01$  & $0.26$ & $35$\ \ \ \ \\
$\sigma [E(B-V)]$              & $0.01$  & $0.05$ & $44$\ \ \ \ \\
\hline
\end{tabular}
\vskip 2mm
\hfill\vfill\parbox[h]{122mm}{\footnotesize 
\underline {Notes:}
The table shows the errors (computed minus model values) of the 
photometric [Fe/H] and $T_{\rm eff}$ for various assumed errors 
in the input parameters. All other input parameters are left at 
their theoretical values, except for the parameter indicated. 
The symbol $\Delta$ in the parameter column means `shifted minus 
theoretical' values. We used Gaussian noise in the random 
simulations (last two rows). In computing the assemble values, 
we used $4\sigma$ clipping. The errors depend on the actual 
parameter regime (see, e.g., Fig.~1). The errors shown in this 
table are average values including the full parameter range.}
\end{table}

For variable stars it is a question of what kind of average 
magnitudes should we use if we need a good approximation of 
the static colors. We use the {\em intensity averaged} magnitudes 
mostly because of the easy availability of these values from the 
OGLE database. Apparently the intensity averages yield closer 
agreement with the static colors, at least for RR~Lyrae stars, 
concerning B, V and K colors (Bono, Caputo \& Stellingwerf 
\cite{bono95}). 

The method was tested on various datasets. For a sample of 42 
Pop.~I dwarfs with spectroscopically determined [Fe/H], $\log g$ 
and $T_{\rm eff}$ values and $BVI$ photometry (e.g., Clementini et al. 
\cite{clem95}; Taylor \cite{tayl03}), we got $\sigma=0.14$~dex 
between the spectroscopic and our photometric [Fe/H]. A much more 
limited sample of 10 Galactic Cepheids yielded $\sigma=0.19$~dex. 
On the set of the 21 Galactic RR~Lyrae stars mentioned above, we 
derived $\sigma=0.15$~dex. No apparent systematic differences 
were observed between the spectroscopic and photometric abundances. 
We conclude that the photometric abundances derived from the $BVI$ 
colors are reasonably accurate on various types of stars. We will 
see that this assessment is confirmed also by the average 
metallicities derived for the MC Cepheids. 

To obtain additional information on the sensitivity of the 
photometric abundances, here we examine the differences obtained 
between the tabulated theoretical values and the ones obtained by 
various perturbations of the corresponding input parameters 
(as given by the stellar atmosphere models). 

In Fig.~1 we show two examples on the effect of changing the 
photometric zero points. We see a dependence on the metallicity 
both in the overall difference and also in the size of scatter 
(due to the various values of $T_{\rm eff}$ and $\log g$). 
On the other hand, it is conceivable that if we are interested in 
{\em relative} abundances, then the method is more accurate, 
especially at higher metallicities. 

To get a closeup on the effects of other possible systematic shifts 
and random errors, we summarize the resulting [Fe/H] errors in Table~1 
(for completeness, we also show $T_{\rm eff}$). As it is expected, 
photometric errors play the most important role, while errors due to 
reddening are smaller. Gravity estimates can be especially erroneous 
without affecting the derived metallicity.

%
%%%%%%%%%%%%%%%%
%  SECTION 3
%%%%%%%%%%%%%%%%
%
\section{[Fe/H] distributions in the MCs}
We applied the above method of [Fe/H] determination to the Cepheid 
database published by Udalski et al. (\cite{udal99}). We used their 
periods, intensity-averaged magnitudes and reddening values as given 
in their $*.tab$ files.\footnote{ftp://sirius.astrouw.edu.pl/ogle/ogle2/var\_stars/lmc/cep/catalog}
\footnote{ftp://sirius.astrouw.edu.pl/ogle/ogle2/var\_stars/smc/cep/catalog/} 
To avoid large errors due to observational noise, as mentioned 
above, we accepted solutions only if they satisfied the 
$\cal D$~$<0.001$ criterion. In addition, we also employed a 
$3\sigma$ clipping to filter out extreme values. The resulting 
[Fe/H] distributions are shown in Fig.~2. 
%
%>>>>>>>>>>
% FIG. 2
%>>>>>>>>>> 
%
\begin{figure}[h]
\includegraphics[width=55mm]{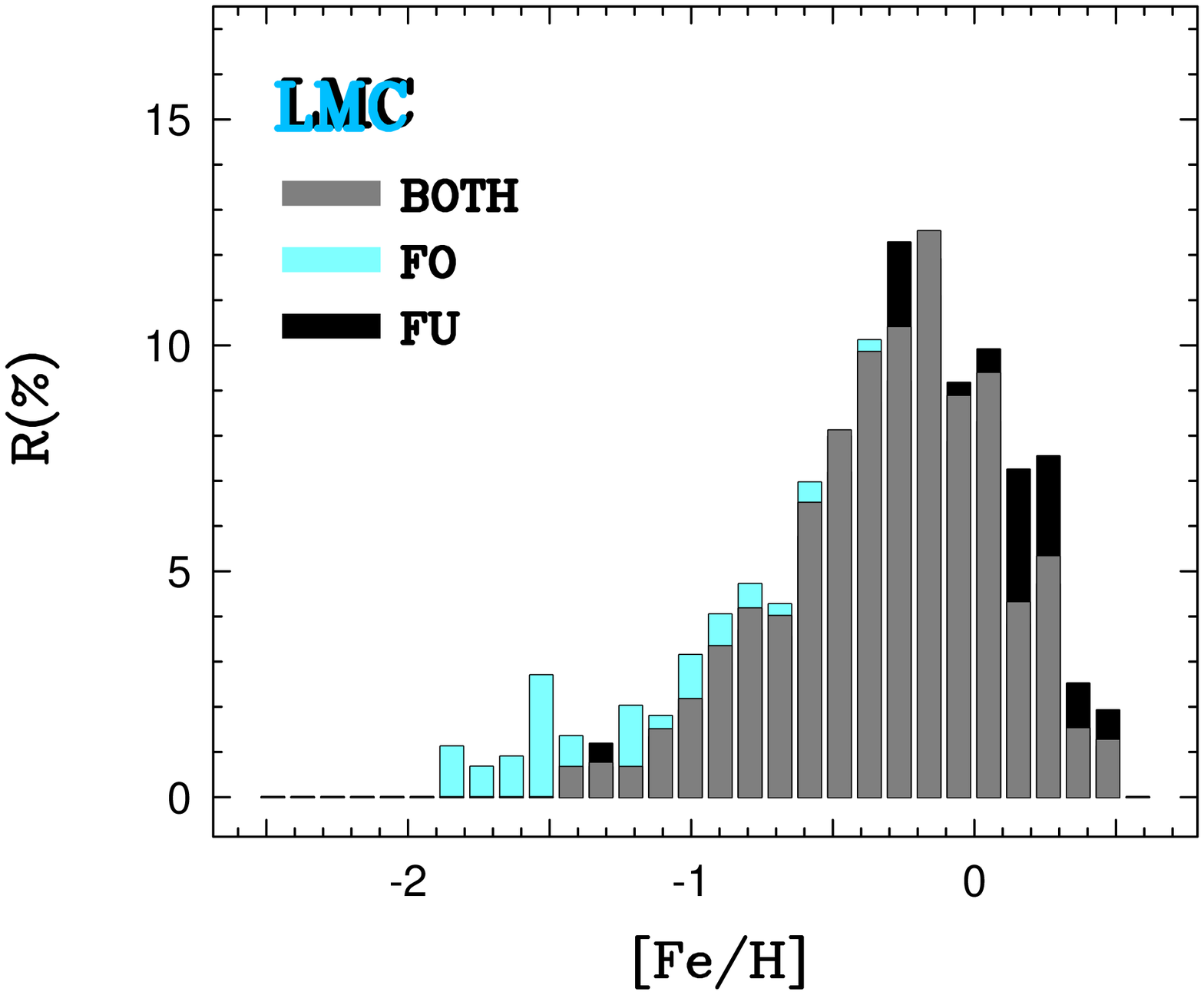}
\hskip 5mm
\includegraphics[width=55mm]{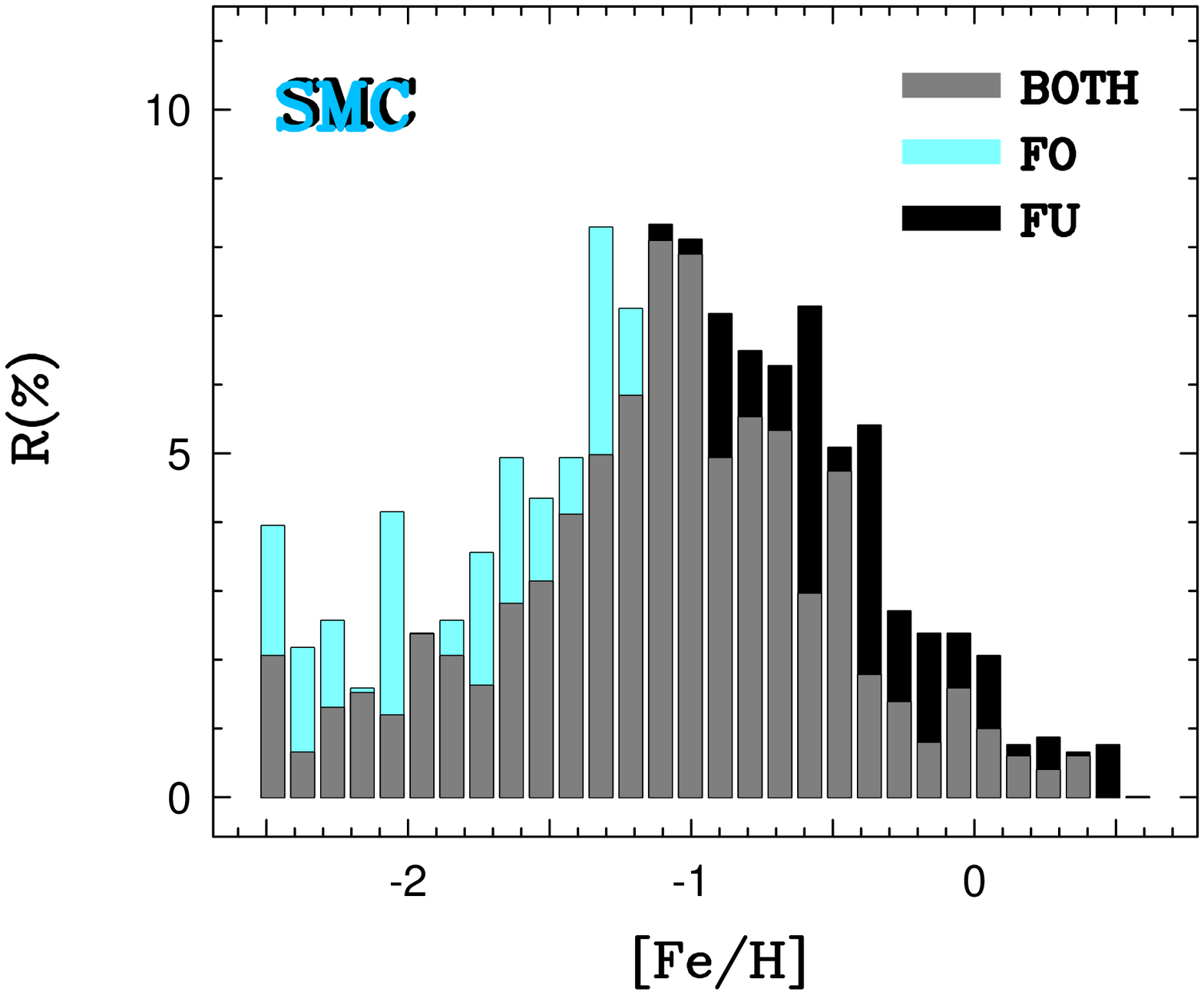}
\caption{Metallicity distributions computed directly from the 
         observed $BVI$ colors.} 
\end{figure}
%

%
%&&&&&&&&&&&&
%  TABLE 2.
%&&&&&&&&&&&&
%
\begin{table}[h]
\caption[]{Average photometric metallicities for the Magellanic Cloud 
Cepheids}
\label{}
\centering
\begin{tabular}{cccccc}
\hline\hline
Cloud    & $N_{\rm FU}$ & [Fe/H]$_{\rm FU}$ & $N_{\rm FO}$  
                        & [Fe/H]$_{\rm FO}$ & 
			[Fe/H]$_{\rm FU}-$[Fe/H]$_{\rm FO}$\\
\hline
LMC      & $676$ & $-0.27\pm0.02$ & $437$ & $-0.43\pm0.02$ & $0.16\pm0.03$ \\
SMC      & $941$ & $-0.97\pm0.02$ & $519$ & $-1.23\pm0.03$ & $0.26\pm0.03$ \\
\hline
\end{tabular}
\end{table}

\noindent
It is clearly seen that there is a surplus of first overtone mode 
variables at the low-metallicity end of the distributions in both 
clouds. The same is true for the fundamental mode variables at the 
high-metallicity end. In both clouds the metallicity spread is large 
but it is clear that most of this comes from observational errors 
(e.g., the spread is larger in the SMC, because of the larger 
photometric errors). When computing the average metallicities, we 
get the result shown in Table~2. Based on the formal errors of the 
mean values, the differences in the overall metallicities seem to 
be significant and are exhibited in the same sense in both clouds. 
We note that: 
(i) the average photometric metallicities are quite close to the 
`canonical' values (e.g., Udalski et al. \cite{udal99}); 
(ii) the derived lower metallicity for the SMC FO variables is the 
same as suggested by Cordier et al. (\cite{cord03}) for the 
low-luminosity Cepheids.     

%
%%%%%%%%%%%%%%%%
%  SECTION 4
%%%%%%%%%%%%%%%%
%
\section{Combination with the PLC relations?} 
In the direct computation of [Fe/H] from the observed color 
indices, large errors are introduced due to the high 
sensitivity to the observational errors. In principle, a 
considerable fraction of this error would be possible to 
eliminate if we could employ the period-luminosity-color 
(PLC) relations and use only the $I$-band magnitudes and 
the period to compute the color indices. Let us assume that 
for each reddening-free magnitude we have the same type of 
relations for {\em all} Cepheids pulsating, e.g., in the 
fundamental mode:   
%
%%%%%%%%%%%%%%
%   EQ. (3)
%%%%%%%%%%%%%%
%
\begin{eqnarray}
W_{\rm B-V} & = & I - 1.86(B-V) = a_0 + a_1\log P_0 \hskip 2mm ,\\
W_{\rm V-I} & = & I - 1.50(V-I) = b_0 + b_1\log P_0 \hskip 2mm ,
\end{eqnarray}
where the $a_i$ and $b_i$ coefficients are independent of the 
metallicity. Similar relations are assumed to be hold also for 
the FO variables. The obvious advantage of the above relations 
is that the errors in the computed color indices become highly 
correlated through the errors of the directly used color $I$. 
Indeed, if we test this method on the stellar atmosphere 
models of Castelli et al. (\cite{cast97}), then, with an error 
of $\sigma_I=0.03$, we get $\sigma_{\rm [Fe/H]}=0.06$ (we recall 
that independent errors of $0.01$ in the color indices cause 
errors of $\sigma_{\rm [Fe/H]}=0.26$ --- see Table~1). We note 
in passing that $T_{\rm eff}$ will not become better estimated. 
With the above noise we get $\sigma_{T_{\rm eff}}=65$. This high 
insensitivity of [Fe/H] against observational noise can be 
understood by the inspection of the linear approximation of 
the color--temperature dependence in the above colors. The 
coefficients of the color terms are nearly cancelled out when 
solved for the metallicity with the aid of the above PLC 
relations (see Kov\'acs \& Walker \cite{kowa99}). 

Although the above method is very promising, there are considerable 
difficulties when looking at the assumptions in more detail. 
First of all, as it follows from the $M$, $L$ and $T_{\rm eff}$ 
dependence of the pulsation period, the PLC relations should have 
some internal scatter, because the mass-luminosity relation is 
not a strict one. Therefore, the derived relation may depend on 
the stellar population studied and cannot be employed in a 
straightforward way to other systems. The population dependence is 
obvious in the case of PL relations and is widely discussed in 
the literature. Although PLC relations have been shown to be much 
less sensitive to such effects, in fine details we might face with 
similar problems as in the case of the PL relations. This question 
has recently been studied by Fiorentino et al. (\cite{fior07}). 
They showed that the PLCs based on the $B$, $V$ colors have 
significant systematic dependence on the metallicity. The effect 
is smaller but not negligible also in the near infrared colors. 
This has led the authors to estimate the metallicities of Galactic 
Cepheids which they showed to exhibit a fine correlation with 
the direct spectroscopic data of Andrievsky and coworkers. 

There are also ambiguities in the derivation of the PLC relations. 
Udalski et al. (\cite{udal99}) employ a $2.5\sigma$-clipping and 
introduce various cuts in their MC sample. In this way they end up 
with the same slope in $W_{\rm V-I}$ for the FU variables in both 
clouds. However, they note that the slopes for the FO variables 
might be different. Also, they did not study $W_{\rm B-V}$, 
presumably due to the paucity of the $B$ data. 

We derived the PLC relations in both unreddened colors as given 
by Eqs. (4.1) and (4.2). We do not apply any period constraint 
on the sample. For each class of variables the first $40$ strongest 
outliers have been omitted iteratively. Table~3 gives the resulting 
regression parameters. By comparing the same types of variables 
in the two clouds, we see that there is a significant difference 
in $b_1$ for the FU variables. Although we use a slightly different 
expression for $W_{\rm V-I}$ than Udalski et al. (\cite{udal99}), 
we think that this result is mainly due to the drop of the period 
constraint. We recall that Udalski et al. (\cite{udal99}) derived 
the same slope in both clouds within fairly narrow error limits. 
On the other hand, we agree with them on the FO variables: the SMC 
variables have a steeper slope in $W_{\rm V-I}$ than the ones in 
the LMC. A similar effect is seen for $W_{\rm B-V}$. Accidentally, 
the FU variables behave in the same way in both clouds in respect 
of the above parameter. It is also important to note that the 
$W_{\rm B-V}$ index yields $\sim 0.15$~mag larger value for the 
relative distance modulus of the clouds than the `canonical' value 
of $0.51$~mag derived from the $W_{\rm V-I}$ index. 

%
%&&&&&&&&&&&&
%  TABLE 3.
%&&&&&&&&&&&&
%
\begin{table}[t]
\caption[]{PLC relations for the Magellanic Clouds Cepheids} 
\label{}
\centering
\begin{tabular}{cccccr}
\hline\hline
\multicolumn{6}{c}{$I - 1.86(B-V) = a_0 + a_1\log P$} \\
Cloud  & Mode & $a_0$  & $a_1$ & $\sigma_{\rm fit}$ & $N$\\
\hline
LMC  &  FU  & $16.021\pm 0.013$ & $-3.590\pm 0.019$ & $0.109$ & $732$  \\
SMC  &  FU  & $16.677\pm 0.008$ & $-3.590\pm 0.016$ & $0.166$ & $1211$ \\
LMC  &  FO  & $15.444\pm 0.008$ & $-3.495\pm 0.023$ & $0.094$ & $469$  \\
SMC  &  FO  & $16.150\pm 0.007$ & $-3.604\pm 0.028$ & $0.150$ & $734$  \\
\hline
\multicolumn{6}{c}{$I - 1.50(V-I) = b_0 + b_1\log P$} \\
Cloud  & Mode & $b_0$  & $b_1$ & $\sigma_{\rm fit}$ & $N$\\
\hline
LMC  &  FU  & $15.915\pm 0.007$ & $-3.314\pm 0.011$ & $0.064$ & $732$  \\
SMC  &  FU  & $16.521\pm 0.006$ & $-3.419\pm 0.012$ & $0.132$ & $1211$ \\
LMC  &  FO  & $15.405\pm 0.006$ & $-3.397\pm 0.015$ & $0.063$ & $469$  \\
SMC  &  FO  & $16.011\pm 0.006$ & $-3.556\pm 0.025$ & $0.134$ & $734$ \\
\hline
\end{tabular}
\end{table}

If we disregard all the above inconsistencies and use the PLC 
method to compute [Fe/H], we get results {\em confirming} the 
metallicity difference between the FO and FU Cepheids in both 
clouds. For the LMC we get average metallicities of $-0.22$ and 
$-0.36$ for the FU and FO variables, respectively. For the SMC 
we get $-1.24$ and $-1.63$, that are considerable lower than 
the ones obtained directly from the color indices (however, the 
relative difference between the FU and FO variables remains 
nearly the same). In both clouds (but especially in SMC) we get 
a considerable decrease in the total metallicity range.

%%%%%%%%%%%%%%%%
%  SECTION 5
%%%%%%%%%%%%%%%%
%
\section{Conclusions}
To lend further support to the hypothesis of systematic metallicity 
spread among the Magellanic Cloud Cepheids, we attempted to derive 
photometric metallicities from the OGLE $BVI$ photometry. Aided by 
the stellar atmosphere models of Castelli et al. (\cite{cast97}) 
we have shown that first overtone variables have systematically 
lower abundances than fundamental ones. The effect is larger in 
the SMC, amounting to $\sim 0.3$~dex (vs. $\sim 0.2$~dex in the 
LMC). This result is in agreement with the ones suggested by 
the evolutionary models of Cordier et al. (\cite{cord03}) on the 
condition of instability strip crossing of low-luminosity Cepheids 
in the SMC. Also, the discrepancy between the pulsational and 
evolutionary ML relations (Beaulieu et al. \cite{beau01}) can be 
largely cured by assuming lower metallicities. Yet another, 
independent result in fitting the periods of the SO/FO Cepheids in
the Magellanic Clouds shows that they require lower metallicities 
than their FO/FU counterparts (Kov\'acs G. \cite{koge06}). We think 
that all these results point toward the reality of the metallicity 
dichotomy between the FO and FU variables. Obviously, a massive 
spectroscopic survey of the Magellanic Clouds Cepheids (similar 
to that of Mottini et al. \cite{mott06}) would be of great value, 
because the photometric abundances are of low accuracy and cannot 
tell the degree of the dichotomy and its more precise relation to 
the luminosity. We also attempted to use the PLC relations to 
derive more accurate color indices and thereby more precise 
abundances. Unfortunately this method remains somewhat doubtful 
due to controversies in the observed relations and due to inherent 
scatter of the theoretical ones (Fiorentino et al. \cite{fior07}).

\section{Acknowledgments}
I would like thank to Robert for the exciting and cheerful years 
I was able to spend at the Physics Department of the University 
of Florida. This long-term collaboration was very pleasant and 
fruitful for me. I feel privileged to be a member of his long 
list of postdoctoral fellows. Thanks are due to Istv\'an D\'ek\'any 
for the careful reading of the manuscript. This work has been 
supported by the Hungarian Scientific Research Foundation (OTKA) 
grant K-60750.

%
%%-----------------------------
%%      your bibliography
%%-----------------------------

%

\end{document}